\documentclass[aps,prl,preprint,floatfix,superscriptaddress,showpacs]{revtex4}

\usepackage{graphicx}%
\usepackage{dcolumn}%
\usepackage{ulem}
\usepackage{soul}

\begin{document}

\title{A new measurement of the antiproton-to-proton flux ratio 
up to 100 GeV 
in the cosmic radiation} 

\author{O. Adriani}
\affiliation{Physics Department of
University of Florence,  
 I-50019 Sesto Fiorentino, Florence, Italy}
\affiliation{INFN, Sezione di Florence,  
 I-50019 Sesto Fiorentino, Florence, Italy}
\author{G. C. Barbarino}
\affiliation{Physics Department of University of
Naples 
``Federico II'',  I-80126 Naples, Italy}
\affiliation{INFN, Sezione di Naples,  I-80126 Naples, Italy}
\author{G. A. Bazilevskaya}
\affiliation{Lebedev Physical Institute, RU-119991
Moscow, Russia}
\author{R. Bellotti}
\affiliation{Physics Department of
University of Bari, I-70126 Bari, Italy}
\affiliation{INFN, Sezione di Bari, I-70126 Bari, Italy}
\author{M. Boezio}
\affiliation{INFN, Sezione di Trieste, I-34012
Trieste, Italy}
\author{E. A. Bogomolov}
\affiliation{Ioffe Physical Technical Institute,  RU-194021 St. 
Petersburg, Russia}
\author{L. Bonechi}
\affiliation{Physics Department of
University of Florence,  
 I-50019 Sesto Fiorentino, Florence, Italy}
\affiliation{INFN, Sezione di Florence,  
 I-50019 Sesto Fiorentino, Florence, Italy}
\author{M. Bongi}
\affiliation{INFN, Sezione di Florence,  
 I-50019 Sesto Fiorentino, Florence, Italy}
\author{V. Bonvicini}
\affiliation{INFN, Sezione di Trieste,  I-34012
Trieste, Italy}
\author{S. Bottai}
\affiliation{INFN, Sezione di Florence,  
 I-50019 Sesto Fiorentino, Florence, Italy}
\author{A. Bruno}
\affiliation{Physics Department of
University of Bari,  I-70126 Bari, Italy}
\affiliation{INFN, Sezione di Bari, I-70126 Bari, Italy}
\author{F. Cafagna}
\affiliation{INFN, Sezione di Bari, I-70126 Bari, Italy}
\author{D. Campana}
\affiliation{INFN, Sezione di Naples,  I-80126 Naples, Italy}
\author{P. Carlson}
\affiliation{Royal Institute of Technology (KTH), Department of Physics, SE-10691 Stockholm, Sweden}
\author{M. Casolino}
\affiliation{INFN, Sezione di Rome ``Tor Vergata'', I-00133 Rome, Italy}
\author{G. Castellini}
\affiliation{ IFAC,  I-50019 Sesto Fiorentino,
Florence, Italy}
\author{M. P. De Pascale}
\affiliation{INFN, Sezione di Rome ``Tor Vergata'', I-00133 Rome, Italy}
\affiliation{Physics Department
of University of Rome ``Tor Vergata'',  I-00133 Rome, Italy}
\author{G. De Rosa}
\affiliation{INFN, Sezione di Naples,  I-80126 Naples, Italy}
\author{D. Fedele}
\affiliation{Physics Department of
University of Florence,  
 I-50019 Sesto Fiorentino, Florence, Italy}
\affiliation{INFN, Sezione di Florence,  
 I-50019 Sesto Fiorentino, Florence, Italy}
\author{A. M. Galper}
\affiliation{Moscow Engineering and Physics Institute,  RU-11540 Moscow, Russia} 
\author{L. Grishantseva}
\affiliation{Moscow Engineering and Physics Institute,  RU-11540 Moscow, Russia} 
\author{P. Hofverberg}
\affiliation{Royal Institute of Technology (KTH), Department of Physics, SE-10691 Stockholm, Sweden}
\author{A. Leonov}
\affiliation{Moscow Engineering and Physics Institute,  RU-11540 Moscow, Russia} 
\author{S. V. Koldashov}
\affiliation{Moscow Engineering and Physics Institute,  RU-11540 Moscow, Russia} 
\author{S. Y. Krutkov}
\affiliation{Ioffe Physical Technical Institute,  RU-194021 St. 
Petersburg, Russia}
\author{A. N. Kvashnin}
\affiliation{Lebedev Physical Institute,  RU-119991
Moscow, Russia}
\author{V. Malvezzi}
\affiliation{INFN, Sezione di Rome ``Tor Vergata'', I-00133 Rome, Italy}
\author{L. Marcelli}
\affiliation{INFN, Sezione di Rome ``Tor Vergata'', I-00133 Rome, Italy}
\author{W. Menn}
\affiliation{Physics Department
of Universit\"{a}t Siegen, D-57068 Siegen, Germany}
\author{V. V. Mikhailov}
\affiliation{Moscow Engineering and Physics Institute,  RU-11540
Moscow, Russia}  
\author{M. Minori}
\affiliation{INFN, Sezione di Rome ``Tor Vergata'', I-00133 Rome, Italy}
\author{E. Mocchiutti}
\affiliation{INFN, Sezione di Trieste,  I-34012
Trieste, Italy}
\author{M. Nagni}
\affiliation{INFN, Sezione di Rome ``Tor Vergata'', I-00133 Rome, Italy}
\author{S. Orsi}
\affiliation{Royal Institute of Technology (KTH), Department of Physics, SE-10691 Stockholm, Sweden}
\author{G. Osteria}
\affiliation{INFN, Sezione di Naples,  I-80126 Naples, Italy}
\author{P. Papini}
\affiliation{INFN, Sezione di Florence,  
 I-50019 Sesto Fiorentino, Florence, Italy}
\author{M. Pearce}
\affiliation{Royal Institute of Technology (KTH), Department of Physics, SE-10691 Stockholm, Sweden}
\author{P. Picozza}
\affiliation{INFN, Sezione di Rome ``Tor Vergata'', I-00133 Rome, Italy}
\affiliation{Physics Department
of University of Rome ``Tor Vergata'',  I-00133 Rome, Italy}
\author{M. Ricci}
\affiliation{INFN, Laboratori Nazionali di Frascati, 
I-00044 Frascati, Italy}
\author{S. B. Ricciarini}
\affiliation{INFN, Sezione di Florence, 
 I-50019 Sesto Fiorentino, Florence, Italy}
\author{M. Simon}
\affiliation{Physics Department
of Universit\"{a}t Siegen, D-57068 Siegen, Germany}
\author{R. Sparvoli}
\affiliation{INFN, Sezione di Rome ``Tor Vergata'', I-00133 Rome, Italy}
\affiliation{Physics Department
of University of Rome ``Tor Vergata'',  I-00133 Rome, Italy}
\author{P. Spillantini}
\affiliation{Physics Department of
University of Florence,  
 I-50019 Sesto Fiorentino, Florence, Italy}
\affiliation{INFN, Sezione di Florence,  
 I-50019 Sesto Fiorentino, Florence, Italy}
\author{Y. I. Stozhkov}
\affiliation{Lebedev Physical Institute,  RU-119991
Moscow, Russia}
\author{E. Taddei}
\affiliation{Physics Department of
University of Florence,  
 I-50019 Sesto Fiorentino, Florence, Italy}
\affiliation{INFN, Sezione di Florence,  
 I-50019 Sesto Fiorentino, Florence, Italy}
\author{A. Vacchi}
\affiliation{INFN, Sezione di Trieste,  I-34012
Trieste, Italy}
\author{E. Vannuccini}
\affiliation{INFN, Sezione di Florence, 
 I-50019 Sesto Fiorentino, Florence, Italy}
\author{G. Vasilyev}
\affiliation{Ioffe Physical Technical Institute, RU-194021 St. 
Petersburg, Russia}
\author{S. A. Voronov}
\affiliation{Moscow Engineering and Physics Institute,  RU-11540
Moscow, Russia}  
\author{Y. T. Yurkin}
\affiliation{Moscow Engineering and Physics Institute,  RU-11540
Moscow, Russia}  
\author{G. Zampa}
\affiliation{INFN, Sezione di Trieste,  I-34012
Trieste, Italy}
\author{N. Zampa}
\affiliation{INFN, Sezione di Trieste,  I-34012
Trieste, Italy}
\author{V. G. Zverev}
\affiliation{Moscow Engineering and Physics Institute,  RU-11540
Moscow, Russia}  

\date{\today}

\begin{abstract}
A new measurement of the cosmic ray antiproton-to-proton flux ratio between 
1 and 100 GeV is presented. The results were obtained with the PAMELA
experiment,   
which was launched into low-earth orbit on-board the Resurs-DK1 satellite on
June 15$^{{\rm th}}$ 2006. 
During 500 days of data collection a
total of about 1000 antiprotons have been 
identified, including 100 above an energy of 20 GeV. The high-energy results
are a ten-fold improvement  
in statistics with respect to all previously published data. 
The data follow the trend expected from secondary production
calculations and significantly constrain 
contributions from exotic sources, e.g. dark matter particle annihilations.
\end{abstract}


\maketitle

Antiprotons can be produced from collisions of energetic
cosmic ray 
particles, primarily protons, with the constituents of the 
interstellar gas such as hydrogen and helium. 
Possible primary sources of galactic antiprotons include 
the annihilation of dark matter
particles~\cite{jun96,ber05} and  
the evaporation
of primordial 
black holes~\cite{haw74,kir81}.
Cosmic ray antiproton experiments can 
probe production and transport properties of cosmic rays
in the galaxy and search for evidence of exotic
production mechanisms. However, such detailed studies of the 
antiproton energy spectrum require measurements with good statistics
over a large energy range. 
Cosmic ray antiprotons were first observed in pioneering experiments
in the 1970s by \citet{bog79} and \citet{gol79} using balloon-borne
magnetic spectrometers. Bogomolov et 
al. observed 2 antiprotons in the kinetic energy range 2-5 GeV while
Golden et al. observed 28 antiprotons in the range 5-12 GeV. Several
other experiments 
followed, covering the kinetic energy range 0.2-50 GeV. 
More than 1000 antiprotons have been observed in the kinetic 
energy range 0.2-4 GeV by the BESS experiment~\cite{yam07} while the
statistics at higher energies is very limited.  
The CAPRICE98~\cite{boe01}, HEAT~\cite{bea01} and MASS91~\cite{hof96}
balloon-borne experiments have observed a total of about 80
antiprotons above 5 GeV. 
However, only two cosmic ray antiprotons with a kinetic energy
above 30 GeV are reported~\cite{boe01}. 

The antiproton-to-proton flux ratio has been measured from 1 to 100
GeV by the PAMELA experiment
(a Payload for
Antimatter Matter Exploration and Light-nuclei Astrophysics), a
satellite-borne apparatus designed to study charged particles in the
cosmic radiation with a particular emphasis on antiparticles. 
The statistics, particularly at high energies, 
is significantly increased compared to the total data sets provided by
all previous experiments. 

The PAMELA apparatus is inserted inside a pressurized container (2~mm
aluminum window) attached
to the Russian Resurs-DK1 satellite and comprises the following
subdetectors:
a time of flight system (ToF);
a magnetic spectrometer;
an anticoincidence system (AC);
an electromagnetic imaging calorimeter;
a shower tail catcher scintillator  and 
a neutron detector.
Technical details about the entire PAMELA instrument and
launch preparations can be found in \cite{pic07}. 

PAMELA has been acquiring data 
since July 11$^{{\rm th}}$ 2006. The
results presented in this letter refer to data acquired in the period
July 2006 to February 2008. More than one billion triggers have been collected
during the total acquisition time of $\sim500$~days.
Events were considered
for further analysis if the reconstructed rigidity exceeded the
vertical geomagnetic cut-off, estimeted using the satellite position, by a factor of 1.3.  
Downward-going particles were selected using the ToF information. The 
time-of-flight resolution of 300~ps ensures that no contamination from albedo particles remains in the 
selected sample. 
The ionization losses (dE/dx) in the ToF scintillators and in the silicon tracker layers
were used to  
select minimum ionizing singly charged particles. 
Furthermore, multiply charged tracks were  
rejected by requiring no spurious signals in the ToF and AC scintillators above the tracking system. 

Particle identification is based on the determination of rigidity by
the spectrometer and the  
properties of the energy deposit and interaction topology in the
calorimeter. 
The analysis technique was validated using the PAMELA
Collaboration's official simulation program tuned using particle beam data.

\begin{figure}[h]
\includegraphics[width=16.8pc]{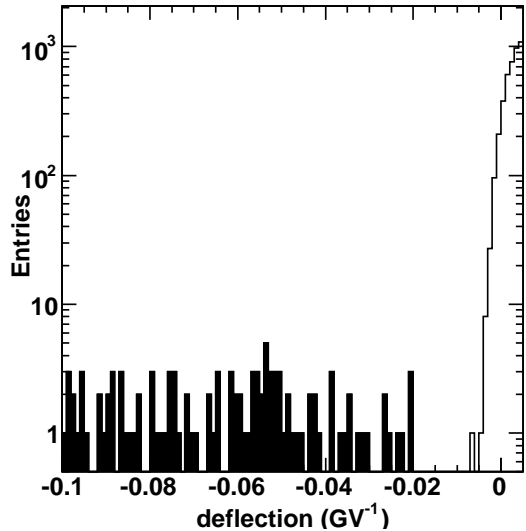}\hspace{2pc}
\caption{The deflection reconstructed by the track
fitting procedure for negatively- and positively-charged 
down-going particles with a reconstructed MDR $\geq$ 850~GeV and that
did not produce an electromagnetic shower in the calorimeter. The
shaded histogram corresponds to the selected antiprotons.  
\label{defl} }   
\end{figure}

The tracking information from the spectrometer is crucial for
selecting antiprotons. 
Due to the finite spectrometer resolution, corresponding to a maximum detectible rigidity (MDR) exceeding 1 TV, high rigidity protons may be assigned the wrong sign of curvature. In
addition 
there is a background from protons that scatter in the 
material of the tracking system and mimic the trajectory of negatively-charged 
particles. In order to accurately measure antiprotons, this
``spillover'' was  eliminated by imposing a set of strict selection
criteria on the quality of the fitted tracks.  
Track fits required the use of at least 4 (3) position 
measurements along the x (y) direction and 
an acceptable $\chi^{2}$ for the fitted track.
To remove spillover protons, clean tracking position 
measurements were required (e.g. no accompanying hits from delta-ray emission) and 
that the MDR, estimated for each event during 
the fitting procedure, should be 10 times larger than the
reconstructed rigidity. 
The deflection (1/rigidity) distribution for positively-
and negatively-charged 
down-going particles, which did not produce an electromagnetic shower
in the calorimeter, is shown in 
Figure~\ref{defl}. 
The sample includes events for which the reconstructed MDR is larger than 850~GV. 
The good separation between negatively-charge particles and spillover
protons is evident. 
As expected, the antiproton tracking requirements limit the 
distribution of spillover protons. 

\begin{figure}[h]
\includegraphics[width=16.8pc]{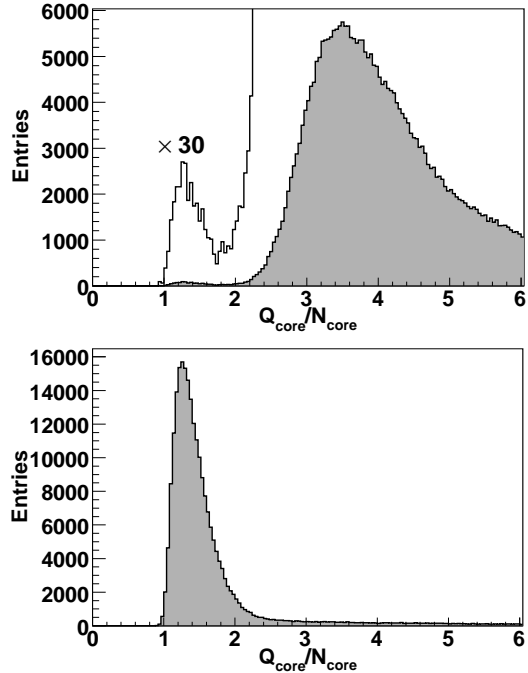}\hspace{2pc}
\caption{An example of a topological calorimeter variable used for antiproton identification (see text for explanation). 
Positively-charged events are shown in the lower plot. The upper plot shows negatively-charged events. 
The vertical scale for the open histogram has been multiplied by a factor of 30 (compared to the filled histogram) for clarity.  
\label{QN} }   
\end{figure}

The calorimeter was used to reject electrons. The longitudinal and transverse 
segmentation of the calorimeter combined with dE/dx measurements from the 
individual silicon strips allow electromagnetic 
showers to be identified with very high accuracy. Using electrons
from simulations and particle beams, and simulated antiprotons,  
we defined an energy dependent calorimeter antiproton
selection~\cite{boe06}. 
Several topological calorimeter variables are used for the antiproton identification.
As an example, the energy density in the shower core weighted by the depth in the calorimeter, 
$Q_{core}/N_{core}$, is shown in Figure~\ref{QN}.
The distribution for the proton-dominated positively-charged sample is peaked at 1.25. In the negatively-charged 
sample, the distribution corresponding to electrons peaks at a higher value, and antiproton events are collected in a 
separate peak positioned similarly to that seen in the positively-charged sample.
The resulting electron contamination was
estimated 
to be negligible across the whole energy range of interest. 
The different, and momentum dependent, interaction
cross sections for 
protons and antiprotons were taken into account estimating the  
calorimeter selection efficiencies as a function of momentum for both
species. These efficiencies  
were studied using both simulated antiprotons and protons, and proton
samples selected from the flight data. 
In the rigidity interval 2 - 100 GV the proton selection efficiency ranges between $0.720 \pm 0.003$ 
and $0.800 \pm 0.012$, whereas the antiproton efficiency ranges between $0.621 \pm 0.003$ and $0.797 \pm 0.012$. 
These
efficiencies were used to rescale the number of selected antiprotons
and protons. 

Possible contamination from pions produced by cosmic-ray interactions
with the PAMELA payload was studied using both simulated and
flight data. 
Both
negatively- and positively-charged pions below 
1~GV were identified using the $\beta$ (velocity) measured by the ToF
system and the 
calorimeter information (to reject electrons and positrons). The
majority of these pion events had hits in the AC 
scintillators and/or large energy deposits in 
one of the top ToF scintillator
 clearly
indicating that they were the product of cosmic ray interactions with
the PAMELA structure or pressure vessel. After
applying all previously described selection criteria, the
energy spectrum of the surviving pions was measured below 1~GV and
compared with 
the corresponding spectrum obtained from simulation
by using
 both GHEISHA and FLUKA
generators~\cite{hof08,bru08}. 
After comparison with the experimental
pion spectrum below 1~GV, a normalization factor for the
simulation, which accounted for all uncertainties related to pion
production and hadronic interactions, was obtained. The normalized
simulated pion spectrum was used to estimate the contamination in the
antiproton sample for rigidities greater than 1~GV. This procedure
resulted in a residual pion contamination of less than 5\% above 2 GV,
decreasing to less than 1\% above 5 GV.
This result was cross-checked between 4 and 8 GV by selecting
antiproton events below the geomagnetic cut-off.  
This sample includes 
re-entrant-albedo~\footnote{Secondary particles produced by cosmic rays
interacting with the 
Earth's atmosphere that are scattered upward but lack sufficient energy
to leave the Earth's magnetic field and re-enter the atmosphere in the
opposite hemisphere but at similar magnetic latitude.}
antiprotons and locally produced pions. 
By scaling the number of such events for the acquisition time 
an upper limit for the
negative pion (and protons with the
wrong sign for the reconstructed deflection) contamination in the
cosmic ray 
antiproton sample was found to be $\sim3\%$, in
agreement with simulations.

\begin{figure}[h]
\includegraphics[width=16.8pc]{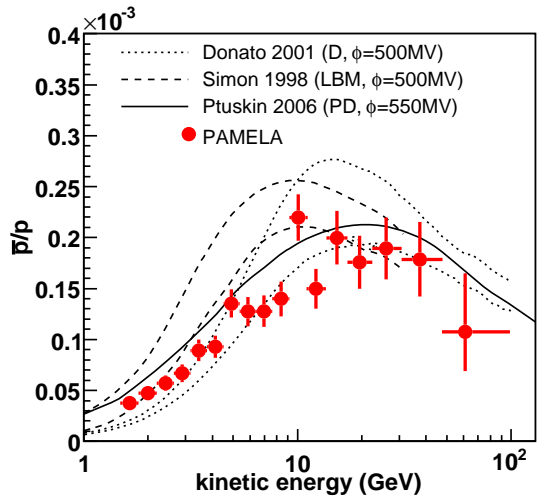}\hspace{2pc}
\caption{The antiproton-to-proton flux ratio 
obtained in 
this work compared with theoretical calculations for a pure secondary
production of antiprotons during the propagation of cosmic rays in the
galaxy. 
The dashed lines
show the upper and lower limits calculated by \citet{sim98} for the
standard Leaky Box Model, while the dotted
lines show the limits from \citet{don01} for a Diffusion model with reacceleration. 
The solid line
shows the calculation by \citet{ptu06} for
the case of a Plain Diffusion model. The curves were
obtained using appropriate solar modulation parameters (indicated as $\phi$) 
for the PAMELA data taking period.
\label{ratio1}}   
\end{figure}

\begin{figure}[h]
\includegraphics[width=16.8pc]{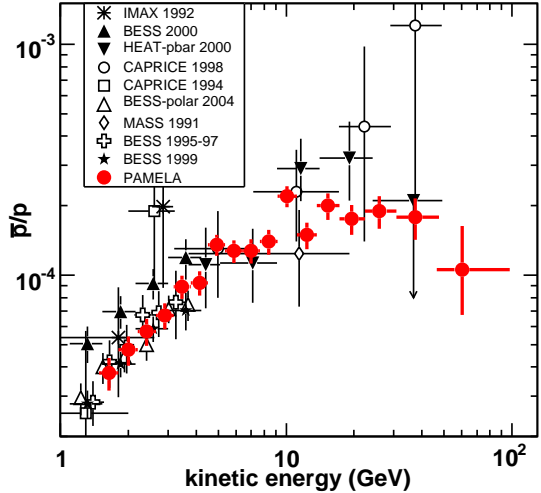}\hspace{2pc}
\caption{The antiproton-to-proton flux ratio obtained in 
this work compared with contemporary measurements 
\protect\cite{hof96,mit96,boe97,boe01,bea01,asa02,ham06}.
\label{ratio2}}   
\end{figure}

Table~\ref{t:pbar} shows the total number of antiprotons and protons that 
survived the data selection.
\begin{table}
\caption{Summary of proton and antiproton 
results. \label{t:pbar}}
\begin{ruledtabular}
\begin{tabular}{ccccc}
Rigidity & Mean Kinetic &\multicolumn{2}{c}{Observed} & Extrapolated \\ 
at & Energy & \multicolumn{2}{c}{number of} &   
\( \frac{\mbox{$\overline{{\rm p}}$ }}{{\mbox
p}} \) \\  
spectrometer &  &
\multicolumn{2}{c}{events} &  \\
GV & GeV & \mbox{$\overline{{\rm p}}$ } & p & at top of payload
 \\ \hline
 2.23 -   2.58 &  1.64 & 39 & 1198039 & $(3.92 \pm 0.63) \times 10^{-5}$ \\
2.58 -   2.99 &  1.99 & 48 & 1144014 & $(4.92 \pm 0.71) \times 10^{-5}$ \\
2.99 -   3.45 &  2.41 & 55 & 1071778 & $(5.91 \pm 0.80) \times 10^{-5}$ \\
 3.45 -   3.99 &  2.89 & 60 & 988666 & $(6.89 \pm 0.89) \times 10^{-5}$ \\
 3.99 -   4.62 &  3.46 & 74 & 903708 & $(9.2 \pm 1.1) \times 10^{-5}$ \\
 4.62 -  5.36 &  4.13 & 71 & 827521 & $(9.6 \pm 1.1) \times 10^{-5}$ \\
 5.36 -  6.23  & 4.91 & 93 & 738028 & $(1.40 \pm 0.14) \times 10^{-4}$ \\
 6.23 -  7.27  & 5.85 & 78 & 653736 & $(1.31 \pm 0.15) \times 10^{-4}$ \\
 7.27 -  8.53  & 6.98 & 69 & 573172 & $(1.32 \pm 0.16) \times 10^{-4}$ \\
 8.53 -  10.1  & 8.37 & 67 & 505503 & $(1.44 \pm 0.18) \times 10^{-4}$ \\
 10.1 -  12.0  & 10.1 & 94 & 449261 & $(2.27 \pm 0.23) \times 10^{-4}$ \\
12.0 -  14.6  & 12.3 & 58 & 405583 & $(1.54 \pm 0.20) \times 10^{-4}$ \\
14.6 -  18.1 & 15.3 & 58 & 301314 & $(2.05 \pm 0.27) \times 10^{-4}$ \\
18.1 -  23.3 & 19.5 & 46 & 270068 & $(1.80 \pm 0.27) \times 10^{-4}$ \\
23.3 -  31.7 & 25.9 & 39 & 211249 & $(1.94 \pm 0.31) \times 10^{-4}$ \\
31.7 -  48.5 & 37.3 & 24 & 136858 & $(1.82 \pm 0.37) \times 10^{-4}$ \\
48.5 - 100.0 & 61.2 & 6 & 57613 & $(1.07^{+0.58}_{-0.39}) \times 10^{-4}$ \\
\end{tabular}
\end{ruledtabular}
\end{table}
The antiproton-to-proton flux ratio was 
corrected for the calorimeter selection efficiencies and for the loss
of particles   
in the instrument itself. It is assumed that 
all antiprotons and protons interacting with the payload material
above and inside the 
tracking system are rejected by the selection criteria. 
The resulting antiproton-to-proton flux ratios 
are given in Table~\ref{t:pbar} and
Figures~\ref{ratio1} and~\ref{ratio2}. The reported errors are
statistical only.  
The contamination was not subtracted from the results and 
should be considered as 
a systematic uncertainty. It is less than a few percent of 
the signal, which is significantly lower than the statistical uncertainty.
Figure~\ref{ratio1} shows the antiproton-to-proton flux ratio measured
by the PAMELA 
experiment compared with  
theoretical calculations assuming pure secondary production 
of antiprotons during the propagation of cosmic rays in the galaxy.  
The PAMELA data are in excellent agreement with recent data from other
experiments, as shown in Figure~\ref{ratio2}.

We have presented the antiproton-to-proton flux ratio over the
most extended energy range ever achieved and we have improved the existing
statistics at high energies by an order of magnitude.
The ratio increases smoothly from about $4 \times 10^{-5}$ at a
kinetic energy of about 1~GeV and levels 
off at about $1 \times 10^{-4}$ for energies above 10~GeV. 
Our results are sufficiently precise to place tight constraints on parameters relevant for secondary 
production calculations, e.g.: the normalization and the index of the diffusion coefficient, the Alfv\'{e}n 
speed, and contribution of a hypothetical ``fresh'' local cosmic ray component \cite{mos03}. Furthermore, an important test criteria for cosmic ray propagation models is their ability to reproduce both the antiproton-to-proton flux ratio and the secondary-to-primary nuclei ratio. Our high energy data (above 10~GeV) places limits on contributions from 
exotic sources, such as dark matter particle annihilations. The antiproton-to-proton flux ratio will be 
modified according to values of the dark matter particle mass, annihilation cross section, and structure 
in the density profile (boost factor).

PAMELA is continuously taking data and the mission is planned to
continue until at least December 2009.
The increase in statistics will allow higher
energies to be studied. 
An analysis for low energy antiprotons (down to $\sim$100~MeV) is in
progress and will be the topic of a future publication \cite{hof08}.

\begin{acknowledgments}
We would like to acknowledge contributions and support from: Italian Space
Agency (ASI), Deutsches Zentrum f\"{u}r Luft- und Raumfahrt (DLR), The
Swedish National Space Board, Swedish Research Council, 
The Russian Space Agency (Roscosmos) and The Russian Foundation for Basic Research.
\end{acknowledgments}

\bibliography{pamela_pbar}

\end{document}